\documentclass[%
 aip,
 jcp,
 amsmath,amssymb,
 reprint,%
]{revtex4-1}

\DeclareRobustCommand*{\citen}[1]{%
  \begingroup
    \romannumeral-`\x 
    \setcitestyle{numbers}%
    \cite{#1}%
  \endgroup   
}

\usepackage{graphicx} 
\usepackage{amsmath} 
\usepackage{xcolor}
\usepackage[separate-uncertainty=true]{siunitx}
\usepackage{amssymb}
\definecolor{steelblue}{RGB}{88,157,200}

\usepackage[
  colorlinks=true,
  linkcolor=steelblue,
  citecolor=steelblue,
  urlcolor=steelblue,
  filecolor=steelblue
]{hyperref}
\usepackage{url}
\usepackage{enumitem}
\usepackage{cleveref}
\crefname{figure}{Fig.}{Figs.}
\Crefname{figure}{Fig.}{Figs.}
\crefname{equation}{Eq.}{Eqs.}
\Crefname{equation}{Eq.}{Eqs.}
\crefname{table}{Tab.}{Tabs.}
\Crefname{table}{Tab.}{Tabs.}
\usepackage{placeins}
\usepackage{upgreek}
\usepackage{xspace}

\newcommand{\mpik}{\affiliation{Max-Planck-Institut f\"ur Kernphysik,
  Saupfercheckweg 1, 69117 Heidelberg, Germany}}
\newcommand{\innsbruck}{\affiliation{Institut f\"ur Ionenphysik und Angewandte
  Physik, Universit\"at Innsbruck, Technikerstra{\ss}e 25, 6020 Innsbruck, Austria}}

\begin{document}

\newcommand*{\addcitation}{\textcolor{blue}}
\newcommand*{\citationneeded}{\textcolor{blue}{[citation needed]}}
\newcommand{\gxe}{GXe\xspace}
\newcommand{\isotope}[2]{\ensuremath{{}^{\mathrm{#1}}\mathrm{#2}}\xspace}

\title{Wavelength-Resolved Timing of Near-Infrared Scintillation in Gaseous Xenon}
\date{October 2026}
\author{Robert Hammann}
\email{hammann@stanford.edu}
\altaffiliation{Present address: Physics Department, Stanford University, Stanford, California 94305, USA}
\mpik
\author{Steffen Form}
\mpik
\author{Humaira Nazneen}
\mpik
\author{Gabriel Sch\"opfer}
\innsbruck
\author{Milan On\v{c}\'ak}
\innsbruck
\author{Teresa Marrod\'an Undagoitia}
\mpik

\begin{abstract}
  The near-infrared (NIR) scintillation signal of xenon has the potential to enhance the performance of xenon-based detectors, but a robust understanding of its properties is a prerequisite for its application.
  In this work, we measure the NIR time response for $\alpha$-particle interactions in gaseous xenon in 12 wavelength bands spanning approximately $(950-1620)$\,nm.
  The time response is decomposed into a fast (ns-scale) component and a dominant slow ($\upmu$s-scale) component.
  The fast-component spectrum coincides with strong atomic emission lines, predominantly 6p$\,\to\,$6s, whereas the slow-component spectrum follows the broad continuum centred around $1.3\,\upmu$m.
  The pressure dependence of the continuum can be described by a two-step kinetic model in which the emitting state is formed by a three-body process and depopulated by a two-body collisional process, with density-independent terms in both rates.
  Interpreting the slower of these as radiative decay yields an intrinsic lifetime of $\sim3\,\upmu$s.
 Theoretical calculations identify a neutral-excimer transition $0_u^- \rightarrow 0_g^-$ and an ionic-dimer transition $1/2_{u} \rightarrow 1/2_g$ as plausible candidates for the continuum emission, with the ionic-dimer transition showing better overall agreement with the measurements.
\end{abstract}

\maketitle

\section{Introduction}
Due to its excellent physical and chemical properties, xenon in gaseous and liquid form is used as a target material in particle detectors.~\cite{Chepel:2012_liquid_noble,Kopec:2023uii}
Xenon is particularly well suited for rare-event searches, including searches for neutrinoless double beta decay~\cite{exo200:2019_final_results,next:2023_limit} and interactions of dark matter particles.~\cite{LZ:2024zvo,XENON:2025vwd,PandaX:2024qfu}
This noble gas is an efficient scintillator with a high vacuum ultraviolet (VUV) light yield, which is the primary signal used in xenon-based detectors.
Gaseous xenon (GXe), however, also exhibits substantial near-infrared (NIR) scintillation.
The NIR timing and yield could provide an additional observable,
complementary to the VUV-light and charge signals, for particle
discrimination in dark matter searches.~\cite{xlzd:2025_design_book,PANDA-X:2024dlo}
If the NIR emission represents additional detectable
quanta, incorporating it could also improve energy resolution, a key
requirement for neutrinoless double-beta-decay searches.~\cite{nexo:2021_sensitivity,
next:2021_sensitivity}
The NIR scintillation signal could also find application in xenon-based calorimeters, such as those used in the MEG\,II~\cite{meg2:2018_design,meg2:2023_calorimeter} or planned for the PIONEER experiment.~\cite{pioneer:2022_snowmass}
To explore this potential, however, the signal needs to be better characterised.
This work focuses on measuring the wavelength-resolved time response and pressure dependence in GXe and comparing the continuum with candidate molecular transitions.

\begin{figure}[t]
  \begin{center}
   \includegraphics[width=\linewidth]{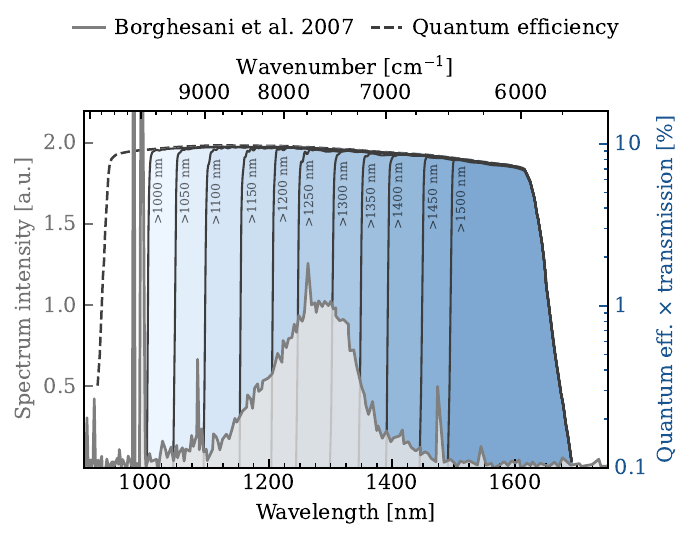}
   \caption{Previously measured near-infrared emission spectrum of gaseous xenon at 1\,bar~\cite{Borghesani:2007} (grey, left axis), with the quantum efficiency of the NIR-sensitive PMT used in this work (dashed line, right axis) and its product with the transmission probability of various long-pass filters (shades of blue) shown in the background.}
   \label{fig:spectrum_qe_filters}
  \end{center}	
\end{figure}

Strong NIR emission in GXe was first observed in the 1980s~\cite{Tanaka:1983_nir,Tanaka:1985_nir} after being predicted approximately a decade earlier.~\cite{Mulliken:1970_xe2}
The dominant emission was associated with a broad continuum centred around \SI{1.3}{\micro m} (\SI{7700}{\per \centi\meter}) and a spectral width of $\sim$\SI{1000}{\per\centi\meter}~\cite{Borghesani:2001} (see \cref{fig:spectrum_qe_filters}).
A red-shift and broadening of this continuum with increasing pressure were subsequently measured and modelled.\cite{Borghesani:2007,Borghesani:2007b,Borghesani:2025}
The emission was attributed to a bound-free transition from the $(3)0_u^+$ state~\footnote{Following the numbering convention of Ref.~\citen{Jonin:2002_xe_potentials_2}.} associated with an atomic 6p-excited state to the dissociative $(1)0_g^+$ state.
Under $\alpha$-particle irradiation, the NIR light yield was measured to be at least 6000\,ph/MeV at 1\,bar,~\cite{Belogurov:2000,Piotter:2023_ir,Hammann:2024_ir} approximately one quarter of the VUV yield.
A monotonic increase of the yield with pressure above \SI{250}{mbar} was reported.~\cite{Tanaka:1983_nir,Piotter:2023_ir}
In addition to the continuum, several narrow emission lines from transitions of excited atomic states exist in the NIR, some of which have already been observed in Ref.~\citen{Lindblom:1988_nir}.
The continuum was found to be strongly suppressed by the addition of O$_2$, but was only weakly affected by SF$_6$, which instead strongly reduced the intensity of the atomic lines.~\cite{Tanaka:1983_nir,Tanaka:1985_nir}

More recent studies have investigated the time response of the NIR emission from $\alpha$-particle interactions.~\cite{Piotter:2023_ir,Hammann:2024_ir}
The dominant part of the emission was found to have a slow $\mathcal{O}(\upmu \mathrm{s})$ time constant, considerably longer than the singlet and triplet decay times in the VUV of $\sim$5\,ns and $\sim$100\,ns.~\cite{Keto:1974_xe_ar_uv,Murayama:2014_time_profile_gxe_lxe,Moutard:1988_pressure_effects}
Only a sub-percent contribution with a faster $\mathcal{O}(\mathrm{ns})$ timing was observed.
The slow component was found to be strongly suppressed by outgassing impurities, suggesting a quenching mechanism.
Additionally, an increase of the effective time constant with decreasing pressure was measured.
However, its microscopic origin remains unclear: neither its association with a particular spectral feature nor the processes governing its formation and decay have been established.
Here, we address these questions by measuring the temporal response in discrete wavelength bands, investigating the formation and decay kinetics of the continuum from its pressure dependence, and evaluating candidate molecular transitions.

\section{Methods}


\subsection{Experimental Setup}
\label{sec:setup}
\begin{figure}[h]
  \begin{center}
   \includegraphics[width=\linewidth]{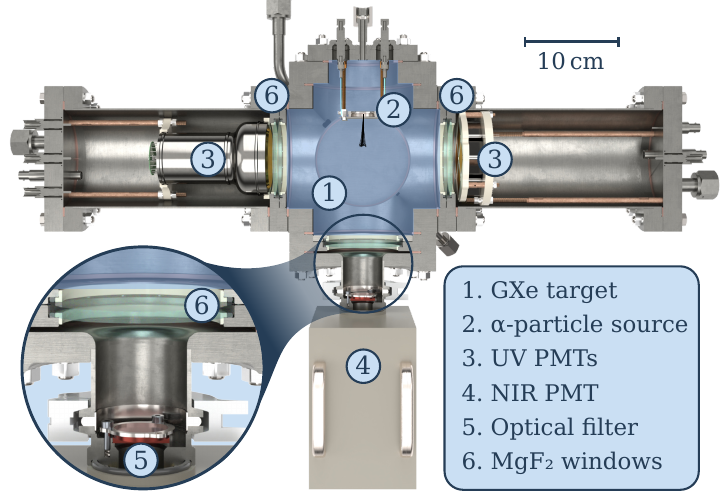}
   \caption{Schematic rendering of the experimental setup. The magnification highlights the optical filter mount in front of the NIR-sensitive PMT. Labelled components are described in the text.}
   \label{fig:setup}
  \end{center}	
\end{figure}

We developed a new setup shown in \cref{fig:setup}, building on the success of the previous apparatus used in Refs.~\citen{Piotter:2023_ir} and \citen{Hammann:2024_ir}.
It consists of a stainless steel cube (CF-100) filled with up to 1.7\,bar of \gxe at room temperature.
The target is irradiated with a \SI{3.7}{kBq} collimated \isotope{241}{Am} $\alpha$-particle calibration source.
Scintillation light is detected by photomultiplier tubes (PMTs) behind three MgF$_2$ windows (5\,mm thick; $>90\,\%$ transmittance between \SI{193}{nm} and \SI{6}{\micro m}; $\sim\,80\,\%$ transmittance at 175\,nm).
The left and right ports hold VUV-sensitive PMTs, while an NIR-sensitive PMT (Hamamatsu H10330C-75) is mounted opposite to the source.
In contrast to the previous configuration, one of the two 3-inch UV PMTs was replaced by an array of four one-inch PMTs.

The new setup includes an optical-filter mount in front of the NIR PMT.
One or two optical filters can be placed in front of the PMT, restricting the detected wavelength range within the sensor’s  \SI{900}{nm}--\SI{1700}{nm} response.
A quick-release mount facilitates switching filters on a short timescale.
We ensured the apparatus was kept light-tight for at least two hours before data acquisition.

As described in Ref.~\citen{Piotter:2023_ir}, the GXe was continuously purified by circulating it through a zirconium hot getter (SAES MonoTorr PS4), which is expected to reduce electronegative impurities to the ppb level.
A new membrane pump allowed continuous recirculation over 17 days with a mass flow of $\sim$(1--2)\,slpm.
The MgF$_2$ windows isolate the \gxe target from the UV photosensors, limiting outgassing into the active volume and helping to maintain stable gas purity.
The measured O$_2$ concentration remained below \SI{25}{ppm}, with the actual concentration expected to be substantially lower.

Events were triggered on the signal in the 3-inch UV-sensitive PMT.
The signals from all six PMTs were recorded with a CAEN V1743 digitizer at a sampling rate of \SI{1.6}{GS/s}.
For each waveform, peaks are identified and quantities such as their arrival time, width, and area (integral under the peak) are computed.
Due to the lower light collection efficiency and quantum efficiency of the NIR PMT compared to the UV PMTs, single-photon NIR pulses are observed while large UV signals with $\sim$1000-photon signals are recorded.
This allows reconstructing the NIR time response by histogramming the arrival time of NIR photons relative to the UV signals from millions of $\alpha$-particle events.~\cite{Piotter:2023_ir}
For each time response, the baseline from dark counts is subtracted and the histograms are normalised to the number of $\alpha$-particle events in each dataset.

\subsection{Theory}
\label{subsec:meth_theory}

In the search for candidate transitions responsible for the broad infrared emission, we calculate excited states of both Xe$_2$ and Xe$_2^+$ and compare the energies, lifetimes, and spectral widths of the associated transitions with the experimental observations.

To model the lowest electronically excited states in Xe$_2$ along the bond elongation, it is necessary to correlate at least 5p, 6s and 6p electrons, while accounting for both static and dynamic correlation. We started with restricted active space self-consistent field (RAS-SCF)~\cite{Malmqvist02:RASSCF,Kreplin19:CASSCF} calculations with the active space comprising 5p, 6s and 6p orbitals of both Xe atoms, i.e., 12 electrons in 14 orbitals, further denoted as (12,14). The number of electrons in the 5p orbitals was limited to 11$-$12. Subsequently, multi-reference configuration interaction (MRCI)~\cite{Werner88:MRCI,Lischka18:MRCI} calculations were performed to obtain more reliable electronic energies, with an analogous restriction of 10$-$12 electrons in 5p orbitals within reference configurations. Finally, spin-orbit calculations were performed employing spin-orbit pseudo-potentials. The arising states were investigated previously in detail in Ref.~\citen{Jonin:2002_xe_potentials_2}.

The electronic states of Xe$_2^+$ were modelled using MRCI with an active space of 11 electrons in 8 orbitals without any restrictions. We considered four electronic states, $^2\Sigma_g^+$, $^2\Sigma_u^+$, $^2\Pi_g$ and $^2\Pi_u$, correlating with the Xe($^1$S) + Xe$^+$($^2$P) asymptote. Accounting for spin-orbit coupling leads to two (1/2)$_g$, two (1/2)$_u$, one (3/2)$_g$ and one (3/2)$_u$ states. \cite{Paidarova:2001} The relative position of curves arising from $J=3/2$ and $J=1/2$ asymptotes of the $^2$P term was shifted by 930 cm$^{-1}$ to match the experimental spin-orbit splitting in the asymptote.

The electronic spectra were modelled using the Franck-Condon analysis for a transition between two bound states and the reflection principle for a transition into an unbound state.~\cite{Oncak10:ReflPrinciple} The vibrational states and wave functions were obtained by numerically solving the Schr\"{o}dinger equation on a grid. For the Franck-Condon analysis, a FWHM of 50 cm$^{-1}$ was assumed for each transition. For the reflection principle, the nuclear distribution in the respective state at a bond length $R$ was modelled as a sum over squared vibrational wave functions in the states $i$, $|\Psi_{\textbf{vib},i}(R)|^2$, employing the Boltzmann distribution at $T$ = 300\,K. The same averaging scheme was used for evaluating the oscillator strength. For modelling the spectral shape in the reflection principle, we used Gaussian convolution with FWHM of 200 cm$^{-1}$. 

The aug-cc-pVQZ atomic orbital basis set was used throughout. All calculations were performed in Molpro, version 2025.3.~\cite{MOLPRO,MOLPRO-WIREs}

\section{Experimental Analysis and Results}

Loose quality cuts are defined to select well-reconstructed $\alpha$-particle events based on the signal distribution across the UV PMTs and the total UV signal area.
Additional cuts on the width and signal area of NIR pulses are applied to reduce electronic noise.

\subsection{Wavelength Scan}
\label{sec:wavelength_scan}
In this work, data were acquired with long-pass filters (Thorlabs) with cut-on wavelengths from \SI{1000}{nm} to \SI{1500}{nm} in steps of \SI{50}{nm}, illustrated in \cref{fig:spectrum_qe_filters}.
With each filter, a dataset of roughly 3 hours was taken.
In addition, the time response without filters was measured (9 hours).
The datasets were recorded in a non-consecutive order of cut-on wavelength to avoid time-dependent artefacts.
All data were acquired at a pressure of \SI{1003\pm5}{mbar}.

\begin{figure}[t!]
  \begin{center}
   \includegraphics[width=\linewidth]{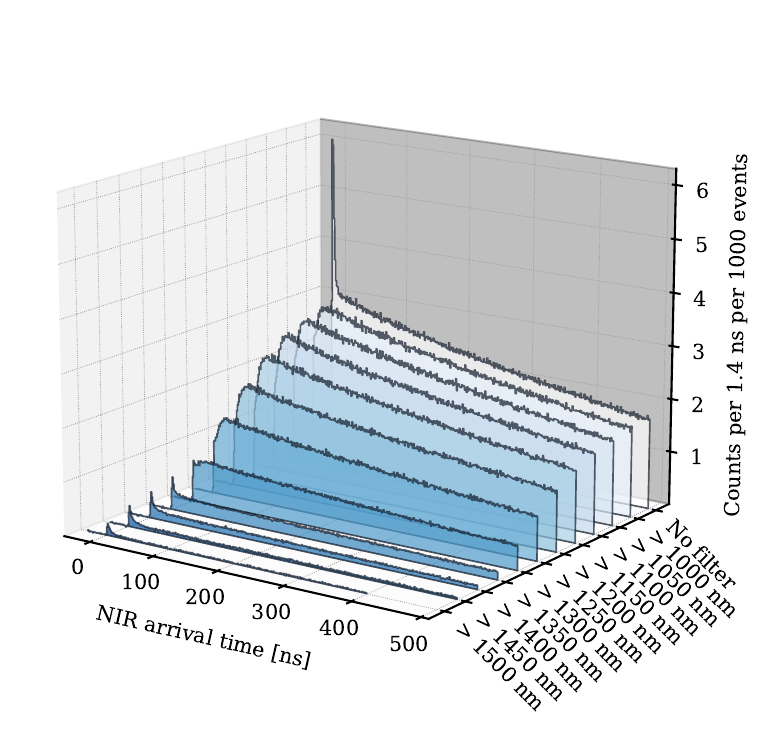}
   \caption{Baseline-subtracted NIR arrival-time distributions measured without optical filter (``No filter'',  white) and with long-pass filters of increasing cut-on wavelength (shades of blue).}
   \label{fig:timing_overlay}
  \end{center}	
\end{figure}

The baseline-subtracted time response for different filters is shown in \cref{fig:timing_overlay}.
The unfiltered distribution agrees well with the shape reported previously observed~\cite{Piotter:2023_ir,Hammann:2024_ir} and can be described by a fast and a slow component.
When the cut-on wavelength exceeds 1000\,nm, thereby rejecting the two strongest atomic lines, the fast component largely vanishes.
For cut-on wavelengths above the continuum peak ($>1350$\,nm), a fast component of smaller amplitude becomes visible.

\begin{figure*}[ht]
  \begin{center}
   \includegraphics[width=0.9\linewidth]{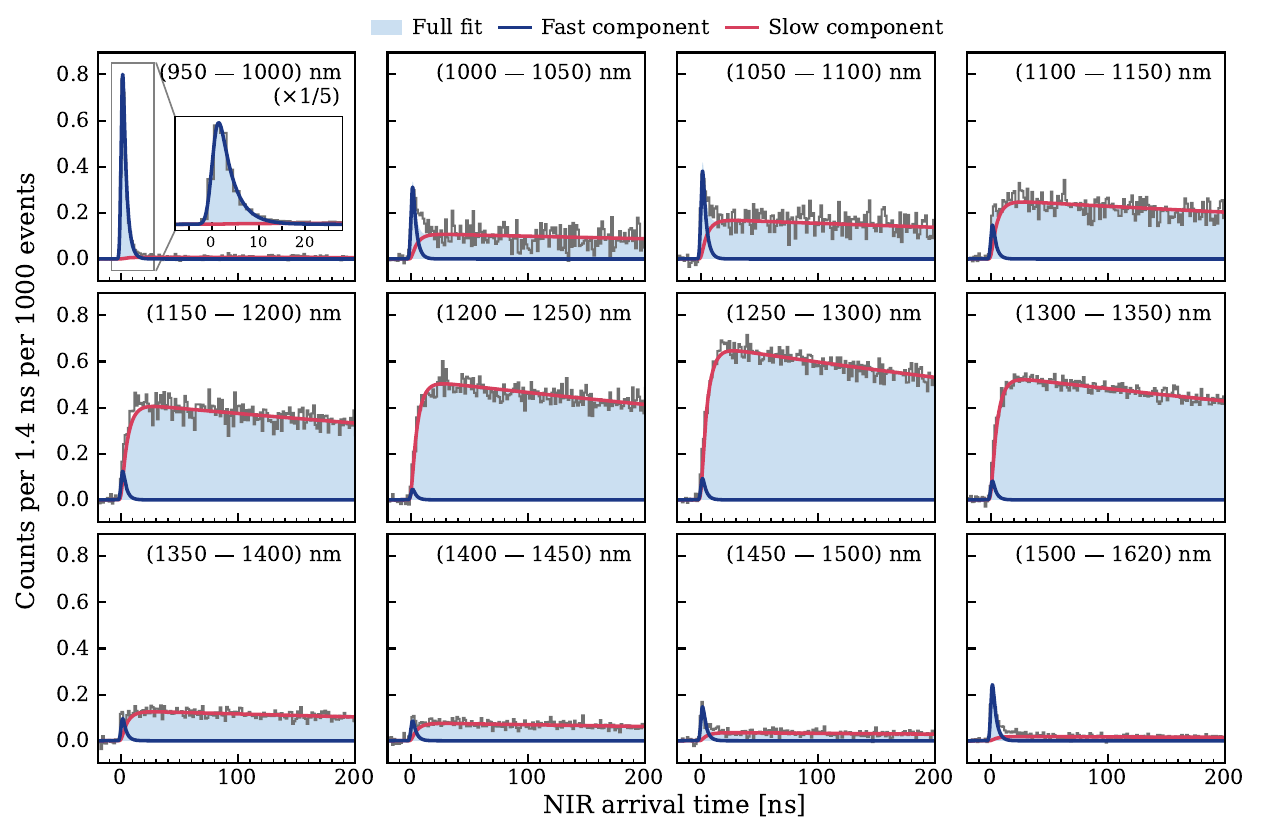}
   \caption{NIR arrival-time distributions in wavelength bands, obtained by subtracting consecutive filter measurements (dark grey). Each panel shows the best-fit model (light blue) decomposed into the fast component (blue) and the slow component (red).  The fast components can be attributed to atomic transition lines, while the slow component is associated with the broad continuum centred around $\sim$1300\,nm (see \cref{fig:spectrum_qe_filters}). The data in the first panel is scaled by $1/5$ and the fast component is magnified in the inset. For visual clarity, the first 200\,ns are shown but the fit is performed in the full time range.}
   \label{fig:timing_wavelngth_band}
  \end{center}	
\end{figure*}

For a more detailed analysis, we subtract measurements from neighbouring long-pass filters to isolate the NIR time response in each wavelength band, after correcting for small differences in filter transmission.
The results are shown in \cref{fig:timing_wavelngth_band}.
Below 1000\,nm, the fast component is most pronounced, suggesting it originates from the two strong 6p$\to$6s atomic lines in this region.~\cite{Borghesani:2007}

The slow component increases steadily in intensity up to 1300\,nm, before decreasing again.
We model the signal time distribution $f$ in each wavelength band for both the fast and slow components as
\begin{equation}
    f(t;\tau_1^\mathrm{eff},\tau_2^\mathrm{eff}) = \frac{\exp(-t/\tau_1^\mathrm{eff}) - \exp(-t/\tau_2^\mathrm{eff})}{(\tau_1^\mathrm{eff}-\tau_2^\mathrm{eff})}\,\Theta(t),
    \label{eq:bateman}
\end{equation}
where $\tau_1^\mathrm{eff}$ and $\tau_2^\mathrm{eff}$ are the effective rise and decay time constants.~\footnote{Since $f(t;\tau_1,\tau_2)$ is invariant
under $\tau_1 \leftrightarrow \tau_2$, we impose
$\tau_1 < \tau_2$ without loss of generality and refer to them
operationally as the rise and decay time constants. This convention
does not uniquely associate either constant with the formation or the de-excitation of the emitting state.}
Their microscopic assignment is discussed below.
This functional form is the solution for consecutive first-order reactions, where one time constant governs the formation of the emitting state and the other its de-excitation.
The sum of the two components is convolved with a Gaussian $G_\sigma$ of width $\sigma$ to account for the finite time resolution.
The full model reads
\begin{align}
    I(t) = &[A_\mathrm{fast}\, f(t;\tau_{\mathrm{fast},1}^\mathrm{eff},\tau_{\mathrm{fast},2}^\mathrm{eff}) \nonumber \\
           &+ A_\mathrm{slow}\,  f(t;\tau_{\mathrm{slow},1}^\mathrm{eff},\tau_{\mathrm{slow},2}^\mathrm{eff})] * G_\sigma(t-t_0),
    \label{eq:fit_function}
\end{align}
where $A_\mathrm{fast}$ and $A_\mathrm{slow}$ are the integrals of the fast and slow components, and $t_0$ is a global time offset accounting for cable delays and trigger effects.
The model in \cref{eq:fit_function} is fitted simultaneously in all 12 wavelength bands using a least-squares fit.
The effective time constants $\tau_{\mathrm{fast},1}^\mathrm{eff}, \tau_{\mathrm{fast},2}^\mathrm{eff}, \tau_{\mathrm{slow},1}^\mathrm{eff}, \tau_{\mathrm{slow},2}^\mathrm{eff}$, the time offset $t_0$, and the resolution $\sigma$ are shared across all bands, while the amplitudes $A_\mathrm{fast}$ and $A_\mathrm{slow}$ are fitted independently for each band.
The best-fit model describes the data well ($\chi^2/n_\mathrm{dof}=3668/4792$).
A slight shoulder visible in the $(1000-1050)$\,nm and $(1500-1620)$\,nm bands is not captured by the fit, which may indicate a third component or a wavelength-dependent time response of the fast component.
The best-fit effective time constants are $\tau_{\mathrm{fast},2}^\mathrm{eff}=(2.96\pm0.07)\,\mathrm{ns}$, $\tau_{\mathrm{slow},1}^\mathrm{eff}=(5.36\pm0.18)\,\mathrm{ns}$, and $\tau_{\mathrm{slow},2}^\mathrm{eff}=(850\pm9)\,\mathrm{ns}$; $\tau_{\mathrm{fast},1}^\mathrm{eff}$ is not resolved.
For the setup-specific nuisance parameters, we find $t_0=(0.18^{+0.03}_{-0.21})\,\mathrm{ns}$ and $\sigma=(1.022	\pm0.018)\,\mathrm{ns}$.

\begin{figure}[h]
  \begin{center}
   \includegraphics[width=\linewidth]{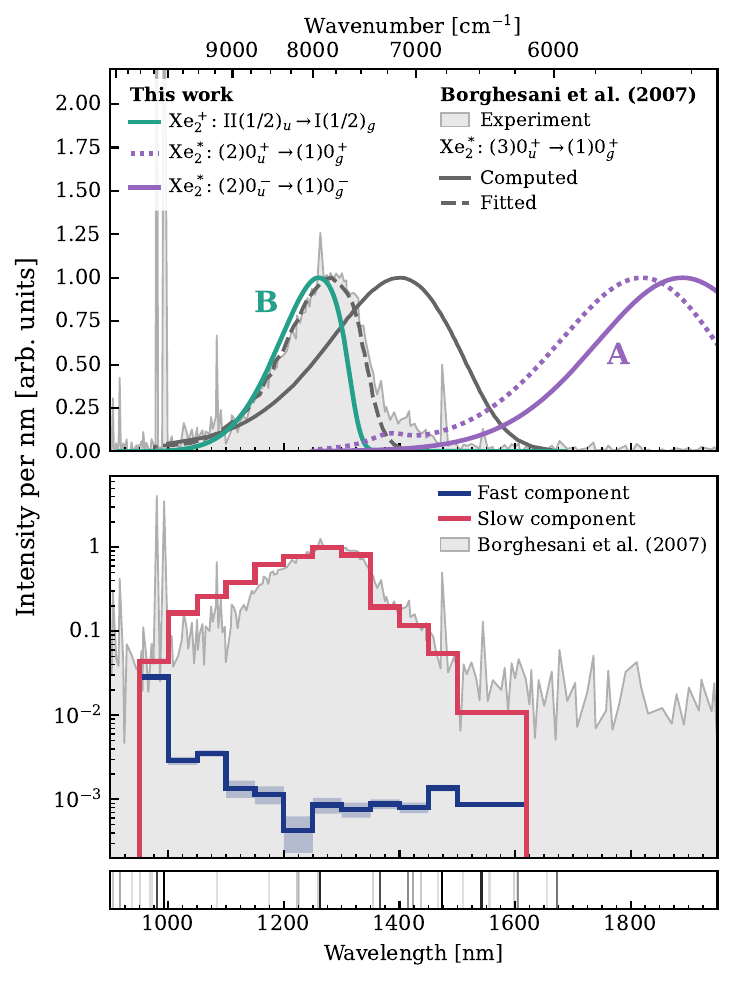}
   \caption{\textit{Top:} The NIR emission spectrum from Ref.~\citen{Borghesani:2007} is shown in light grey, together with the excimer continuum computed by the authors of Ref.~\citen{Borghesani:2007}  for the (3)$0_u^+ \rightarrow (1)0_g^+$ transition using the potentials of Ref.~\citen{Jonin:2002_xe_potentials_2} (solid  grey), and a variant in which the lower-state potential was modified to fit the data (dashed grey). The continua computed in this work from the excimer ($(2)0_u^+\to (1)0_g^+$ and $(2)0_u^-\to (1)0_g^-$) and ionic dimer (II$(1/2)_u\to$I$(1/2)_g$) transitions  are shown as purple dotted, purple solid, and green solid lines, respectively.
   \textit{Middle:} Integrals of the slow (red) and fast (blue) fit components in 50\,nm wavelength bands, shown as a coarsely binned spectrum. The grey reference spectrum is scaled to allow easy comparison with the spectrum obtained for the slow component. \textit{Bottom:} Positions of strong atomic transition lines,~\cite{nist:2005_spectroscopic_data} with darker lines corresponding to stronger transitions.}
   \label{fig:spectrum_comparison}
  \end{center}	
\end{figure}

The integrals $A_\mathrm{fast}$ and $A_\mathrm{slow}$ -- corrected for the average quantum efficiency in the wavelength band -- are shown in \cref{fig:spectrum_comparison}, providing a coarsely-binned spectrum of the fast and slow components.
The slow component reproduces the shape and location of the broad continuum around $1.3\,\upmu$m, consistent with the spectra previously reported.~\cite{Tanaka:1983_nir,Tanaka:1985_nir,Borghesani:2007,Borghesani:2007b,Borghesani:2025}
The fast component peaks below 1000\,nm and higher intensities coincide with bins containing strong catalogued atomic transition lines.~\cite{nist:2005_spectroscopic_data}
One exception is the $(1000-1050)$\,nm band, which contains no known strong atomic lines but might receive contributions from the strong line just below 1000\,nm leaking through the filter.

\subsection{Dependence on Pressure}
\begin{figure*}[ht]
  \begin{center}
   \includegraphics[width=0.9\linewidth]{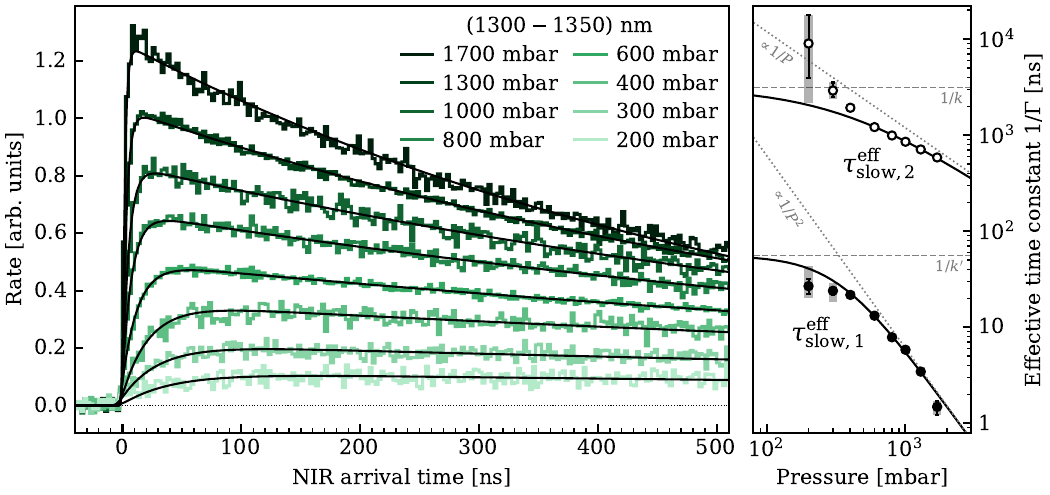}
   \caption{Pressure dependence of the NIR signal in the $(1300-1350)$\,nm wavelength band, which is dominated by the broad continuum emission. \textit{Left:} Baseline-subtracted arrival-time distributions at various pressures (shades of green), each scaled for visual clarity, with the best-fit four-parameter kinetic model overlaid (solid black lines). \textit{Right:} Pressure dependence of the fitted rise and decay time constants, $\tau_\mathrm{slow,1}^\mathrm{eff}$ (closed circles) and $\tau_\mathrm{slow,2}^\mathrm{eff}$ (open circles), both decreasing with increasing pressure. Error bars indicate statistical uncertainties and shaded bands the total uncertainty, computed as the quadrature sum of statistical and systematic uncertainties. Solid lines show $1/\Gamma_1$ and $1/\Gamma_2$ of the four-parameter model (\cref{tab:rate_coefficients}), dashed horizontal lines the density-independent limits $1/k'$ and $1/k$, and dotted lines the pressure-dependent terms only.}
   \label{fig:spectrum_comparison_pressure}
  \end{center}	
\end{figure*}

To gain insight into the underlying kinetics of the continuum emission, we study the pressure dependence of its time response in the $(1300-1350)$\,nm band.
The band is selected to avoid contamination from strong atomic transition lines using one long-pass and one short-pass filter placed in front of the NIR-sensitive PMT.
The pressure is varied over the range $(200-1700)$\,mbar, and the results are shown in \cref{fig:spectrum_comparison_pressure}.

In a first step, we fit the time response at each pressure independently with \cref{eq:fit_function}, setting $A_\mathrm{fast}=0$ and leaving the two effective time constants $\tau_\mathrm{slow,1}^\mathrm{eff}$ and $\tau_\mathrm{slow,2}^\mathrm{eff}$ free (\cref{fig:spectrum_comparison_pressure}).
The nuisance parameters $t_0$ and $\sigma$ are loosely constrained using pressure-dependent measurements of the VUV time response and the fast NIR component.
Variants with one or both constraints removed are used to estimate systematic uncertainties.
Above $\sim$\,600\,mbar the shorter constant approximately scales as $\tau_\mathrm{slow,1}^\mathrm{eff} \propto P^{-2}$ and the longer one, $\tau_\mathrm{slow,2}^\mathrm{eff}$,  decreases slightly slower than $\propto P^{-1}$.
Below $\sim$\,400\,mbar $\tau_\mathrm{slow,1}^\mathrm{eff}$ seems to level off and $\tau_\mathrm{slow,2}^\mathrm{eff}$ becomes poorly constrained as it grows well beyond the recorded time window of 550\,ns.

As noted in \cref{sec:wavelength_scan}, the difference of two exponentials in \cref{eq:bateman} is the solution of a two-step consecutive process: a precursor, populated promptly by the $\alpha$-particle interaction, forms the emitting state, which subsequently decays.
Writing the effective time constants as rates, $\Gamma_i = 1/\tau_\mathrm{slow,\mathit{i}}^\mathrm{eff}$, the time response reads
\begin{equation}
    I(t) = A\left[\,\frac{e^{-\Gamma_2 t}-e^{-\Gamma_1 t}}{1/\Gamma_2-1/\Gamma_1} \,\Theta(t) \right]\ast G_\sigma(t-t_0)\,.
    \label{eq:kinetic_model}
\end{equation}
Each rate is the sum of all processes that depopulate the respective state, and its dependence on the xenon number density $n=P/(k_\mathrm{B}T)$ at $T = 300$\,K reflects the reaction order: density independent for unimolecular processes such as radiative decay, $\propto n$ for two-body and $\propto n^2$ for three-body collisions.
Guided by the observed scaling, we parametrise
\begin{align}
\Gamma_1 =&\, k_3 n^2 + k'\label{eq:gamma1}\\
\Gamma_2 =&\, k_2 n + k.\label{eq:gamma2}
\end{align}

A simultaneous fit of \cref{eq:kinetic_model} at all pressures, keeping $A$ free at each pressure and $t_0$, $\sigma$ constrained as above, describes the data with four instead of sixteen rate parameters (\cref{tab:rate_coefficients}), at a cost of $\Delta\chi^2 = 50.5$ for 12 fewer degrees of freedom relative to the independent fits.
Removing either of the two density-independent terms increases $\chi^2$ by 277 ($k' = 0$) or 456 ($k = 0$) for one parameter, so both constant terms are required.
\begin{table}[h]
  \centering
  \caption{Best-fit rate coefficients of the four-parameter model, $\Gamma_1 = k_3 n^2 + k^\prime$ and $\Gamma_2 = k_2 n + k$, from a simultaneous fit at all pressures. Systematic uncertainties are the envelope of fits with the constraints on $t_0$ and $\sigma$ removed individually and together.}\label{tab:rate_coefficients}
  \begin{tabular}{lcl}
    \hline\hline
    Parameter & Value & Unit \\
    \hline
    $k_3$ & $2.82 \pm 0.07\,(\mathrm{stat})\,{}^{+0.14}_{-0.06}\,(\mathrm{syst})$ & $10^{-31}\,\mathrm{cm^6\,s^{-1}}$ \\
    $k_2$ & $3.47 \pm 0.05\,(\mathrm{stat})\,{}^{+0.01}_{-0.03}\,(\mathrm{syst})$ & $10^{-14}\,\mathrm{cm^3\,s^{-1}}$ \\
    $1/k'$ & $56^{+5}_{-4}\,(\mathrm{stat})\,{}^{+21}_{-5}\,(\mathrm{syst})$ & ns \\
    $1/k$ & $3140^{+160}_{-140}\,(\mathrm{stat})\,{}^{+20}_{-90}\,(\mathrm{syst})$ & ns \\
    \hline\hline
  \end{tabular}
\end{table}
The best-fit models are overlaid in \cref{fig:spectrum_comparison_pressure} and describe the data well ($\chi^2/n_\mathrm{dof} = 2122/1756$), with some apparent mismodelling near the peak at the highest pressure.
Note that the pulse amplitudes are scaled arbitrarily for visual clarity.
A quantitative analysis of the pressure-dependent light yield, including corrections for the pressure-dependent light collection efficiency, is provided in Ref.~\citen{Piotter:2023_ir}.

\section{Theoretical Results}
\label{sec:theory}

In our modelling, we considered transitions in both Xe$_2$ and Xe$_2^+$ as possible sources of the experimental broad emission continuum. Properties of the calculated bands are collected in \cref{tab:exc_prop}. For Xe$_2$, we considered the lowest-lying bound states corresponding to the $^2$P$_{1/2}$\,6s and $^2$P$_{1/2,3/2}$\,6p asymptotes for every irreducible representation that lie at least 5000\,cm$^{-1}$ above the first excited state of Xe$_2$ as we expect these states to be long-lived and possibly provide signals in the experimentally covered range. In this approximation, energetically higher-lying states are expected to relax readily to lower-lying electronic states through conical intersections (crossing of states) before they reach a state where they have sufficiently long lifetimes to be observed experimentally. The dissociation asymptotes are well reproduced and the low-lying potentials are consistent with literature,~\cite{Jonin:2002_xe_potentials_1,Jonin:2002_xe_potentials_2} with quantitative shifts that reflect different computational approaches.

Out of the considered initial states, several irreducible representations can be excluded a priori: The lowest-lying bound states of 1$_{g/u}$ and 2$_{g/u}$ cross with dissociative states of the same irreducible representations, and are thus not expected to live long enough on the relevant experimental timescale.
The lowest 3$_g$ state is also unlikely to be long-lived because of its limited stability.~\cite{Jonin:2002_xe_potentials_2}
The lowest 3$_u$ state is comparatively stable and could be considered for fluorescence transitions, although relatively high in energy.~\cite{Jonin:2002_xe_potentials_2}
However, as this state arises from 5d orbitals that are not included in our computational approach, we do not model the respective transitions here. For the remaining states, i.e., (2)$0_g^+$, (2)$0_g^-$, (2)$0_u^+$, (2)$0_u^-$, we analyse all possible transitions to lower-lying states with considerable oscillator strength (above 0.01).

\begin{table}[h]
    \centering
    \caption{Comparison of properties of calculated transitions and the observed continuum: emission maximum $\tilde{\nu}_{\mathrm{max}}$, FWHM (both in cm$^{-1}$) and lifetime $\tau$ (in ns). The candidate transitions A and B are indicated. For the two transitions to the $(1)1_u$ state, the intrinsic Franck-Condon width is narrow, so the quoted FWHM is dominated by the empirical broadening (see \cref{fig:xe2_spectra_si} in the supplementary material).}
    \begin{tabular}{l l c c c}
        \toprule
         & Transition & $\tilde{\nu}_{\mathrm{max}}$ & FWHM & $\tau$ \\
         \hline
          &Measured & 7700 & 1000 & 3140\footnotemark[1]\\
        \hline
         &Xe$_2$, $(2)0_g^+\to (1)1_u$ & 9200 & $<200$ & 70 \\
          &Xe$_2$, $(2)0_g^-\to (1)1_u$ & 9000 & $<200$ & 70 \\
          &Xe$_2$, $(2)0_u^+\to (1)0_g^+$ & 5400 & 1000 & 300\footnotemark[2] \\
         \textbf{A} &Xe$_2$, $(2)0_u^-\to (1)0_g^-$ & 5200 & 1000 & 200 \\
         \textbf{B} &Xe$_2^+$, II$(1/2)_u\to$I$(1/2)_g$ & 7900\footnotemark[3] & 800 & 600\footnotemark[3] \\
         \hline
\multicolumn{5}{l}{\emph{Competing transition into the ground state}}\\
&Xe$_2$, $(2)0_u^+\to (0)0_g^+$ & 71000 & 4300 & 2 \\
        \toprule
    \end{tabular}
    \label{tab:exc_prop}
\footnotetext[1]{Value of $1/k$ (\cref{tab:rate_coefficients}), interpreting $k$ as the radiative decay rate; the reverse assignment would correspond to $1/k' \approx 56$\,ns.}
\footnotetext[2]{The observed time constant would be dominated by the fast competing transition into the ground state.}
\footnotetext[3]{A shift of 930\,cm$^{-1}$ was used to obtain this value (see Methods\,\cref{subsec:meth_theory}).}
\end{table}

For (2)$0_g^{+/-}$ states, the lowest-lying bound states can relax into the bound (1)$1_u$ state. These transitions lie at about 9000\,cm$^{-1}$, with oscillator strengths of about 0.26, corresponding to lifetimes of about 70\,ns. Their spectra are however too narrow (see~\cref{tab:exc_prop} and \cref{fig:xe2_spectra_si} in the supplementary material) compared to the experiment due to their Franck-Condon character. 

Bound-free transitions from the $(2)0_u^{+/-}$ states to the dissociative $(1)0_g^{+/-}$ states are analogous to the previously proposed $(3)0_u^+\to(1)0_g^+$ transition.~\cite{Borghesani:2007}
We note that the (3)$0_u^+$ state is metastable in our calculations, and we thus do not model the respective spectrum.
The spectrum calculated for this transition in Ref.~\citen{Borghesani:2007}, using the potential-energy curves from Ref.~\citen{Jonin:2002_xe_potentials_2}, is redshifted relative to the measured continuum, as shown in \cref{fig:spectrum_comparison}.
In that work, the lower-state potential was subsequently adjusted empirically to reproduce the observed spectrum.
The spectra of the (2)$0_u^{+/-}\,\rightarrow\,(1)0_g^{+/-}$ are shown in \cref{fig:spectrum_comparison}. They have FWHM values of about 1000\,cm$^{-1}$ and thermally averaged oscillator strengths of 0.17 and 0.31, corresponding to lifetimes of 300 and 200\,ns, and are thus possible candidates for the observed continuum. Their position is considerably below the experimental signal by about 2500\,cm$^{-1}$, although we note that the computational accuracy is limited for such highly excited electronic states.

\begin{figure}
    \centering
    \includegraphics[width=\linewidth]{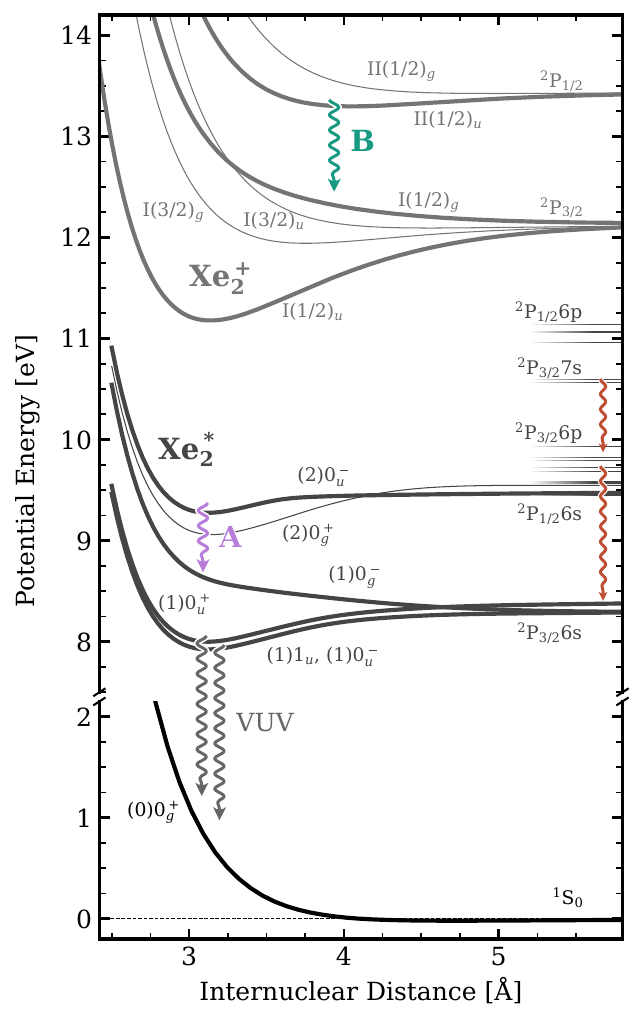}
    \caption{Selection of potential energy curves for Xe$_2^+$ and Xe$_2^*$ computed in this work. The two candidate transitions for the \SI{1.3}{\micro m} continuum are indicated by A and B. In addition, atomic NIR transitions between 7s, 6p, and 6s states (Ref.~\citen{nist:2005_spectroscopic_data}) are sketched, and the VUV continuum to the ground state is shown.}
    \label{fig:potential_diagram_illustration}
\end{figure}

For the (2)$0_u^+ \rightarrow$ (1)$0_g^+$ transition considered here (see \cref{fig:potential_diagram_illustration}) and the (3)$0_u^+ \rightarrow$ (1)$0_g^+$ transition suggested previously,~\cite{Borghesani:2007} an additional argument disfavours their assignment to the observed band: (2)$0_u^+$ and (3)$0_u^+$ states may also fluoresce into the ground state (0)$0_g^+$ with oscillator strengths of 0.22 and 0.16 calculated at bond lengths of $3.2$\,\AA{} and $3.6$\,\AA, respectively. This corresponds to the considerably shorter lifetimes of about $(1-2)$\,ns due to the large energy of these transitions above 70\,000\,cm$^{-1}$.
Decay to the ground state would therefore dominate over the competing NIR transition, with fewer than one percent of decays from the $(2)0_u^+$ and $(3)0_u^+$ states expected to produce a NIR photon.
Because all radiative channels depopulate the same upper state, the NIR emission would inherit this nanosecond-scale lifetime.
Both the small branching fraction and the short lifetime therefore disfavour these states as the origin of the intense NIR continuum.
Note that transitions from a $0_u^-$ state do not suffer from this problem as the transition into the (0)$0_g^+$ ground state is forbidden.

In Xe$_2^+$, we considered only the lowest six states corresponding to the $^2$P$_{3/2,1/2}$ asymptotic states of Xe$_2^+$. Among these, the only bound state that could potentially contribute to signals at $\sim$\,1300\,nm is II(1/2)$_u$ (see \cref{fig:potential_diagram_illustration}). From this state, four target states are available for fluorescence. Out of these, transitions from II(1/2)$_u$ into I(1/2)$_u$ and I(3/2)$_u$ are electric-dipole forbidden by parity ($u\!\rightarrow u$). The transition II(1/2)$_u$ $\rightarrow$ I(3/2)$_g$ is calculated to have a negligible oscillator strength (below $10^{-5}$ at $R=4$\,\AA) and is thus of no importance on the experimental timescale. However, the transition $\mathrm{II(1/2)}_u\rightarrow\mathrm{I(1/2)}_g$ has a thermally averaged oscillator strength of 0.038, corresponding to a lifetime of about 600\,ns. The modelled spectrum of the transition has a FWHM of 800\,cm$^{-1}$ and is shown in \cref{fig:spectrum_comparison}.

\section{Discussion}
\subsection{Spectral origin of the fast and slow components}
The observed fast component coincides with the positions of strong atomic transition lines, while the slow component reproduces the shape and location of the broad continuum around $1.3\,\upmu$m, consistent with the spectrum reported previously.~\cite{Tanaka:1983_nir,Tanaka:1985_nir,Borghesani:2007,Borghesani:2007b,Borghesani:2025}
We therefore infer that the fast component originates from atomic transitions and the slow component from the continuum.
This is further supported by the strong purity dependence of the slow component reported in Ref.~\citen{Piotter:2023_ir} and of the continuum in Ref.~\citen{Borghesani:2025}.
The slow component is well described by the same effective rise and decay time constants across the full wavelength range, consistent with a common primary emission process resulting in the continuum.
Its spectrum reproduces the xenon NIR continuum previously measured with a different apparatus, photosensor, and detector materials, strongly disfavouring material fluorescence as a dominant contribution.

The individual atomic transitions likely have different lifetimes, but the wavelength-binned data are consistent with all of them contributing on the nanosecond timescale.
Their individual time profiles cannot be distinguished at the present temporal and spectral resolution.
Resolving line-specific lifetimes would require a finer wavelength selection than achievable with the present setup.

\subsection{Formation and decay kinetics of the continuum}
We measured the pressure dependence of the continuum's time profile in a band without strong atomic lines.
Both $\tau_\mathrm{slow,1}^\mathrm{eff}$ and $\tau_\mathrm{slow,2}^\mathrm{eff}$ decrease by roughly one order of magnitude between 200\,mbar and 1700\,mbar, demonstrating a strong pressure dependence of the processes involved in the continuum emission.
The observed pressure dependence of the effective parameters motivates the kinetic model of \cref{eq:kinetic_model,eq:gamma1,eq:gamma2}, which describes the time profiles at all pressures with four instead of sixteen rate parameters, at a moderate increase in $\chi^2$.

The $n^2$ term in $\Gamma_1$ is characteristic of a three-body process.
The fitted coefficient, $k_3 = 2.8\times10^{-31}\,\mathrm{cm^6\,s^{-1}}$, is comparable to the reported reaction coefficients at 300\,K for both $\mathrm{Xe^+ + 2Xe\to Xe_2^+ + Xe}$, $(1.5-2.0)\times10^{-31}\,\mathrm{cm^6\,s^{-1}}$, and $\mathrm{Xe^* + 2Xe\to Xe_2^* + Xe}$, $(0.4-1.4)\times10^{-31}\,\mathrm{cm^6\,s^{-1}}$.~\cite{Vitols:1973,Neves:2010,Leichner:1976_first_continuum,Bonifield:1980_xenon_rate_constants,Moutard:1988_pressure_effects}
We therefore interpret $\Gamma_1$ as the depopulation rate of the precursor and its $n^2$ term as the three-body formation of the emitting molecule, although $k_3$ alone does not discriminate between an ionic and a neutral precursor.
Since \cref{eq:kinetic_model} is symmetric in $\Gamma_1$ and $\Gamma_2$, the reverse assignment cannot be excluded from the time profiles alone, but it would require a three-body depopulation of the emitting state whose coefficient happens to coincide with known association rates.

In our interpretation, $k'$ describes a density-independent process with $1/k' \approx 56$\,ns, which dominates over the three-body formation below $\sim$\,330\,mbar.
The time profile cannot distinguish whether this process competes with the formation or itself populates the emitting state, since the two cases differ only in the amplitude of \cref{eq:kinetic_model}, which is not analysed here, and not in its shape.
We consider a competing loss more natural: in this case, the fraction of precursors that form the emitting state, $k_3n^2/(k_3n^2+k')$, falls below one half for $P<330\,$mbar, so that the yield readily drops towards low pressure, as reported in Refs.~\citen{Tanaka:1983_nir,Piotter:2023_ir}.
One possibility for its origin is radiative decay of an excited atomic precursor into states that do not feed the emitting state.

Correspondingly, $\Gamma_2$ is assigned to the depopulation of the emitting state, and $k$ its radiative decay rate with an intrinsic lifetime of $1/k \approx 3.1\,\upmu$s.
The $k_2n$ term, with $k_2\simeq3.5\times10^{-14}\,\mathrm{cm^3\,s^{-1}}$, could represent two-body collisional de-excitation from the emitting state to a lower-lying state, as proposed in Ref.~\citen{Bonifield:1980_xenon_rate_constants}.
The measured constant is about four orders of magnitude smaller than the gas-kinetic collision rate coefficient ($\sim2\times10^{-10}\,\mathrm{cm^3\,s^{-1}}$), corresponding to a reaction probability of only $\sim10^{-4}$ per collision.
Two caveats apply to the stated value of $1/k \approx 3.1\,\upmu$s.
First, any density-independent non-radiative loss, for example quenching by impurities at constant partial pressure, adds to $k$, so $1/k$ is strictly a lower limit on the radiative lifetime.
Second, $1/k$ exceeds the 550\,ns recording window by a factor of about six.
It is therefore mainly determined indirectly from the deviation of $\tau_\mathrm{slow,2}^\mathrm{eff}$ from a pure $1/P$ scaling at higher pressures rather than directly from the convergence at  lower pressures.

\subsection{Previously proposed transitions}
Two transitions have previously been proposed as the origin of the continuum.
From the appearance of the continuum in Ar-Xe mixtures and its absence in mixtures with other noble gases, the authors of Refs.~\citen{Tanaka:1983_nir,Tanaka:1985_nir} suggested that a highly excited excimer state correlated to the 9s asymptote, decaying to a dissociative state correlated to 6p, could be responsible.
Such a highly excited state would need to remain long-lived despite the many state crossings in this energy region. 
We therefore regard this assignment as unlikely, although it is not tested directly by the present calculations.

A second suggestion is the transition from the (3)$0_u^+$ bound state,~\cite{Borghesani:2007} motivated by its relation to the 6p states, which are clearly populated given the strong 6p\,$\to$\,6s emission lines observed in the spectrum.
Three arguments disfavour this assignment.
First, as discussed in \cref{sec:theory}, the $(3)0_u^+$ state decays to the ground state within $(1-2)$\,ns, so that fewer than one percent of its decays would produce a NIR photon, which is difficult to reconcile with a NIR yield of about one quarter of the VUV yield.~\cite{Hammann:2024_ir}
Second, the NIR emission would inherit this ns-scale lifetime, which is incompatible with both density-independent constants, $1/k' \approx 56$\,ns and $1/k \approx 3.1\,\upmu$s, irrespective of which of them is assigned to the radiative decay.
Third, the simultaneous disappearance of the atomic lines and persistence of the continuum with SF$_6$ reported in Refs.~\citen{Tanaka:1983_nir,Tanaka:1985_nir} disfavours a mechanism in which the observed 6p population directly feeds the continuum.
The same arguments apply to the $(2)0_u^+ \to (1)0_g^+$ transition computed in this work.

\subsection{Candidate transitions identified in this work}
Our calculations identify two transitions that avoid the fast ground-state channel and produce a continuum near the measured one: the ionic-dimer transition II$(1/2)_u\to$I$(1/2)_g$ (transition B in \cref{fig:potential_diagram_illustration}) and the neutral-excimer transition $(2)0_u^-\to(1)0_g^-$ (transition A).

\textit{Transition A (neutral excimer).}
Because the $(2)0_u^-$ state has negative parity, electric-dipole decay to the $(0)0_g^+$ ground state is forbidden, avoiding the fast competing channel of the $0_u^+$ candidates. The computed spectrum has the measured width but lies $\sim$\,2500\,cm$^{-1}$ below the measured continuum, although the computational accuracy is limited, as mentioned above.
Its precursor would be a directly excited atom, which forms the excimer by three-body association.\footnote{%
Population via electron-ion recombination is disfavoured: the continuum is essentially unaffected by the electron scavenger SF$_6$,~\cite{Tanaka:1983_nir,Tanaka:1985_nir,Hunter:1989_sf6} recombination is reported to become negligible towards 1\,bar, with no electric-field dependence observed there,~\cite{Mimura:2010_recombination,Suzuki:1982_time_dependence,Leardini:2021_time_and_band} and the slow NIR component persists at intermediate electric fields.~\cite{Hammann:2024_ir}}
Because SF$_6$ strongly suppresses the observed atomic-line emission,~\cite{Tanaka:1983_nir,Tanaka:1985_nir} excimer formation, which takes $1/(k_3n^2)\approx6$\,ns at 1\,bar, would have to outcompete this quenching for the continuum to persist.
In the proposed kinetic interpretation, the computed lifetime of 200\,ns is a factor of $\sim$\,15 shorter than $1/k$.

\textit{Transition B (ionic dimer).}
The computed spectrum reproduces the position ($7900$ vs.\ $7700\,\mathrm{cm^{-1}}$), width ($800$ vs.\ $1000\,\mathrm{cm^{-1}}$) and shape of the measured continuum.
The emitting state could be formed by three-body association of a Xe$^+$ ion.
The computed radiative lifetime of 600\,ns is a factor of $\sim$\,5 shorter than $1/k \approx 3.1\,\upmu$s.
We regard this as a plausible but not compelling agreement, given that the uncertainty of the computed oscillator strength cannot be quantified.
The two-body term $k_2 n$ could then correspond to an inefficient collisional relaxation of II$(1/2)_u$ to lower-lying ionic states.
In this picture, charge transfer provides a possible explanation for the reported impurity dependence.
The ionization energy of O$_2$ (approximately $12.1$\,eV~\cite{NISTWebBook:IonizationEnergy}) lies about $1.1$ eV below the calculated energy of the emitting $\mathrm{II}(1/2)_u$ state (approximately $13.2$\,eV), making process $\mathrm{Xe}_2^{+}[\mathrm{II}(1/2)_u]+\mathrm{O}_2\to2\mathrm{Xe}+\mathrm{O}_2^+$ energetically plausible, providing a possible explanation for why O$_2$ quenches the continuum.
By contrast, the ionization energy of SF$_6$ (approximately $15.3$\,eV~\cite{NISTWebBook:IonizationEnergy}) lies above the calculated state energy, so the same simple charge-transfer channel is not accessible, consistent with the absence of strong quenching.

Taken together, the ionic-dimer transition provides the more consistent description of the data, matching the spectral position and width, with a computed lifetime within a factor of about five of $1/k$ and a plausible qualitative explanation for the reported O$_2$ and SF$_6$ dependence, even though the neutral-excimer transition A cannot be excluded.
Other ionic species not included in the present calculations may also warrant consideration, including excited $\mathrm{Xe}_3^+$ ionic trimers (clusters; see e.g. Ref.~\citen{Helm:1976_xe3plus}).
In argon, the analogous $\mathrm{Ar}_3^+$ species has been proposed as the terminal emitter in a cascade underlying the third continuum.~\cite{Wieser:2000_third_continuum}

A firm assignment will require improved calculations and additional data to better constrain the microphysics of the continuum emission.
An extended acquisition window would improve the low-pressure determination of $\tau_\mathrm{slow,2}^\mathrm{eff}$, while a broader pressure range would assess the validity of the model.
Measurements with a controlled admixture of different impurities would provide additional constraints on the underlying processes.

\section{Conclusions}
We measured the time response of the NIR scintillation of gaseous xenon in discrete 50\,nm wavelength bands between 950\,nm and 1620\,nm.
The response comprises a fast component whose wavelength dependence coincides with strong atomic emission lines and a dominant slow component whose spectrum reproduces the broad continuum near $1.3\,\upmu$m.
The slow component is described by a common pair of time constants across the measured wavelength range, consistent with a common dominant emission channel.
In the continuum-dominated $(1300-1350)$\,nm band, both effective time constants decrease by approximately one order of magnitude between 200 and 1700\,mbar, but with different density dependences.
A two-step kinetic model with four rate constants describes the time profiles at all pressures.
The faster rate contains a three-body term, $k_3 = 2.8\times10^{-31}\,\mathrm{cm^6\,s^{-1}}$, of the same order as the association rates of both the ionic dimer and neutral excimers, and the slower rate a two-body term, $k_2 = 3.5\times10^{-14}\,\mathrm{cm^3\,s^{-1}}$.
Both rates contain a density-independent term.
Under the preferred assignment, interpreting the slower of these as the radiative decay of the emitting state gives an intrinsic lifetime of about $3\,\upmu$s.

Theoretical calculations identify a neutral-excimer transition, $(2)0_u^-\to(1)0_g^-$, and an ionic-dimer transition, II$(1/2)_u\to$I$(1/2)_g$, as candidates for the continuum.
The ionic dimer reproduces the position, width and shape of the continuum, its computed lifetime is of the same order as the measured $\upmu$s-scale constant, and it offers a plausible explanation for the reported impurity dependence.
It therefore provides the more consistent description, although the neutral-excimer transition cannot be excluded on the basis of the present data.
Identifying the transition responsible for the continuum, and establishing how the NIR scintillation signal can be used to improve the performance of xenon-based detectors, particularly in the liquid phase,~\cite{Hammann:2025_ir_tpc} remain goals of ongoing work.

\section*{Acknowledgements}
\noindent
We thank Roberto Santorelli, Edgar S\'anchez Garc\'ia, and Vicente Pesudo for their help with 
the design of the MgF$_2$ chamber windows, Marjolein Troost for her insightful comments on this manuscript, and Viviane Schmidt for fruitful discussions and for 
connecting the experimental and theoretical authors of this work. The computational results presented here have been achieved using the LEO HPC infrastructure of the University of Innsbruck. The project is supported by the Max Planck Society.

\section*{Data Availability}
\noindent
The data presented in this work as well as the computed potentials and spectra are available online.~\cite{zenodo_data_release}


\section*{References}
\bibliography{references}

\section*{Supplementary Material}
\setcounter{figure}{0}
\renewcommand{\thefigure}{S\arabic{figure}}
\renewcommand{\theHfigure}{S\arabic{figure}}

\begin{figure*}[h!]
  \begin{center}
   \includegraphics[width=1.0\linewidth]{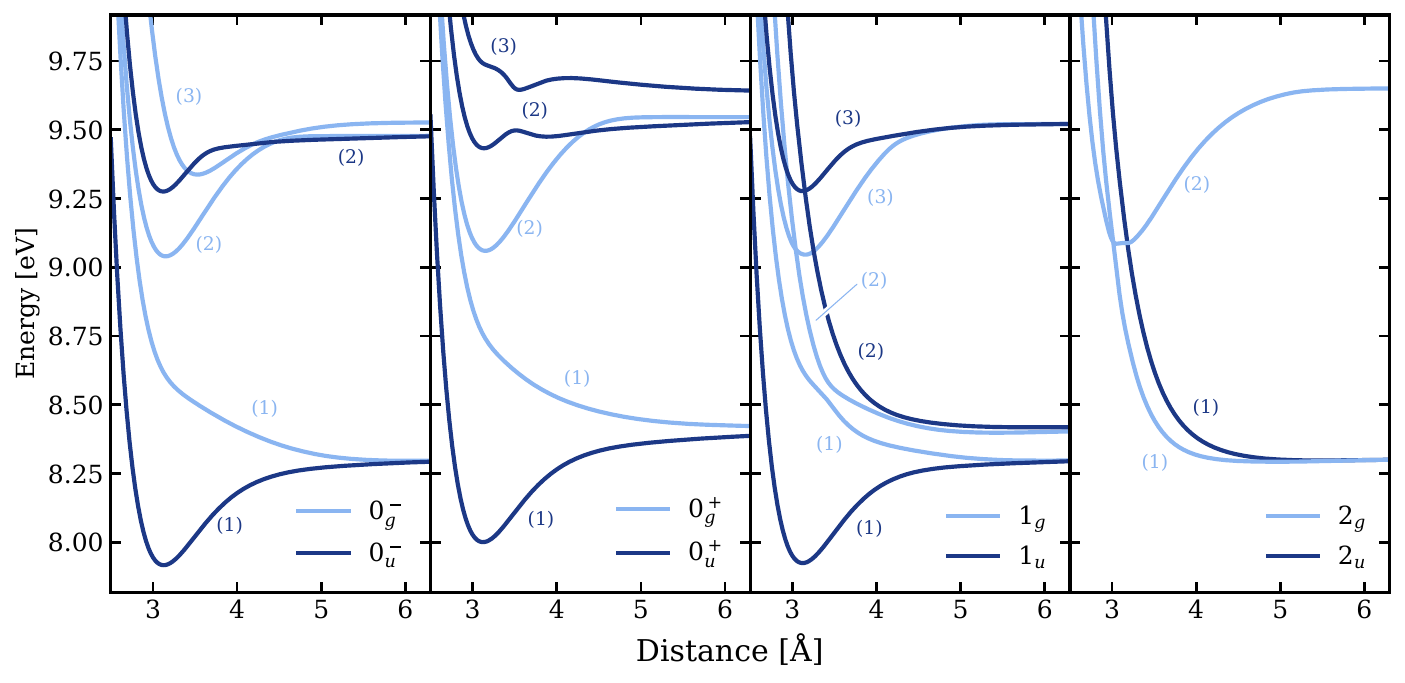}
   \caption{Potential energy curves for Xe$_2^*$ computed in this work.}
   \label{fig:xe2_potentials_si}
  \end{center}	
\end{figure*}

\begin{figure*}[h!]
  \begin{center}
   \includegraphics[width=1.0\linewidth]{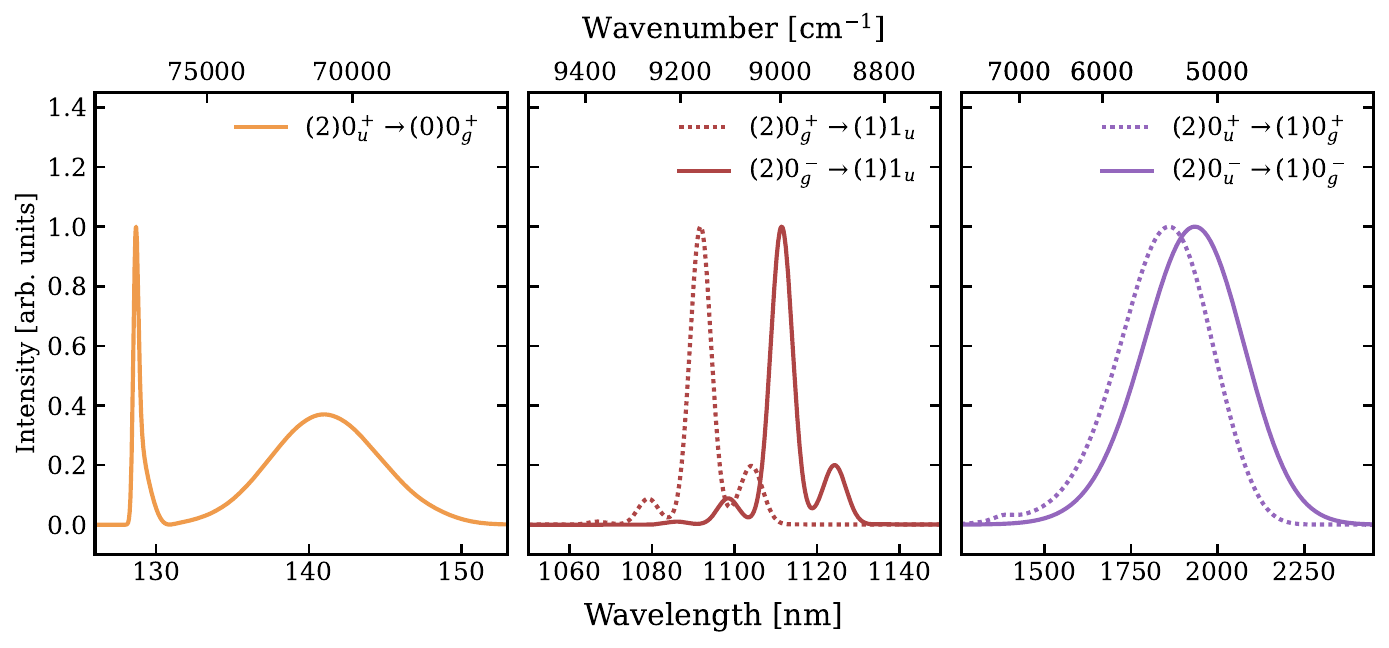}
   \caption{Emission spectra computed in this work.}
   \label{fig:xe2_spectra_si}
  \end{center}	
\end{figure*}

\Cref{fig:xe2_potentials_si,fig:xe2_spectra_si} show the potential energy curves and emission spectra for Xe$_2^*$ computed in this work.

\end{document}